\documentclass[backref, twocolumn, breaklinks, colorlinks]{aastex63}   
\hypersetup{linkcolor=blue,citecolor=blue,filecolor=cyan,urlcolor=magenta}

\usepackage{enumitem}
\usepackage{mathtools}
\usepackage{graphicx}
\usepackage{amssymb}
\usepackage{amsmath}
\usepackage{ulem}
\usepackage{bm}
\usepackage{epstopdf}
\usepackage{color}
\usepackage[caption=false]{subfig}

\begin{document}
\tighten

\title{{Constraining the Emission Geometry and Mass of the White Dwarf Pulsar AR Sco using the Rotating Vector Model}}

\author{Louis du Plessis}
\affil{Centre for Space Research, North-West University, Private Bag X6001, Potchefstroom 2520, South Africa}

\author{Zorawar Wadiasingh}
\affil{Astrophysics Science Division, NASA Goddard Space Flight Center, Greenbelt, MD 20771, USA}
\affil{Centre for Space Research, North-West University, Private Bag X6001, Potchefstroom 2520, South Africa}

\author{Christo Venter}
\affil{Centre for Space Research, North-West University, Private Bag X6001, Potchefstroom 2520, South Africa}

\author{Alice K. Harding}
\affil{Astrophysics Science Division, NASA Goddard Space Flight Center, Greenbelt, MD 20771, USA}

\shorttitle{AR Sco RVM}

\begin{abstract} 
\noindent We apply  the standard radio pulsar rotating vector model to the white dwarf pulsar AR Sco's optical polarization position angle swings folded at the white dwarf's spin period as obtained by \citet{Buckley2017}. Owing to the long duty cycle of spin pulsations with a good signal-to-noise ratio over the entire spin phase, in contrast to neutron star radio pulsars, we find well-constrained values for the magnetic obliquity $\alpha$ and observer viewing direction $\zeta$ with respect to the spin axis. We find $\cos\alpha=0.060^{+0.050}_{-0.053}$ and $\cos\zeta=0.49^{+0.09}_{-0.08}$, implying an orthogonal rotator with an observer angle $\zeta={60.4^\circ}^{+5.3^\circ}_{-6.0^\circ}$. This orthogonal nature of the rotator is consistent with the optical light curve consisting of two pulses per spin period, separated by $180^\circ$ in phase. Under the assumption that $\zeta \approx i$, where $i$ is the orbital inclination, and a Roche-lobe-filling companion M star, we obtain $m_{\rm WD} = 1.00^{+0.16}_{-0.10} M_\odot$ for the white dwarf mass. These polarization modeling results suggest the { that non-thermal emission arises from a dipolar white dwarf magnetosphere and close to the star,} with synchrotron radiation (if non-zero pitch angles can be maintained) being the plausible loss mechanism, marking AR Sco as an exceptional system for future theoretical and observational study.
\end{abstract} 

\section{Introduction} \label{intro}

AR Scorpii (AR Sco) is an intriguing binary system containing a putative white dwarf (WD) with an M-dwarf companion. The system exhibits pulsed non-thermal radio, optical, and X-ray emission, likely of synchrotron radiation (SR) origin \citep{Marsh2016,Buckley2017,Takata2018}. AR Sco has a binary orbital period of $P_{\rm b} =3.56$ hours, a WD spin period of $P= 1.95$ minutes, and a beat period of 1.97 minutes. The light cylinder radius\footnote{This is the radius where the corotation speed equals that of light in vacuum.} and orbital separation are $R_{\rm{LC}}=c/\Omega\sim 6\times 10^{11}$~cm and $a\sim 8\times 10^{10}$~cm, respectively, with $\Omega$ the spin angular frequency. Thus, the M star is located within the WD's magnetosphere at $a \approx 0.13 R_{\rm LC}$. A change in spin period $\dot{P}=3.9\times10^{-13}\,\rm{s \, s^{-1}}$ was inferred by \cite{Marsh2016} but was disputed by \cite{Potter2018} who argued that the observations were too sparse to derive an accurate spin-down time scale for the WD pulsar. With more extensive observations, \cite{Stiller2018} firmly established $\dot{P} = 7.18\times10^{-13}\,\rm{\, s \, s^{-1}}$ that is almost twice as large as the value reported in \cite{Marsh2016}, but consistent with the constraints in \cite{Potter2018}. The concomitant spin-down power is $\dot{E}_{\rm rot} = I_{\rm WD} \Omega \dot{\Omega} \approx 5 \times 10^{33}$ erg s$^{-1}$ for a fiducial WD moment of inertia $ I_{\rm WD} = 3 \times 10^{50}$ g cm${}^2$.  A lack of Doppler-broadened emission lines from accreting gas suggests the absence of an accretion disk\footnote{ This is supported by the fact that the X-ray luminosity is only 4\% of the total, optically-dominated observed luminosity and is only $\sim1\%$ of the X-ray luminosities of typical intermediate polars, and also from the fact that all optical and ultraviolet emission lines originate from the irradiated face of the M-dwarf companion.}.  More recently, X-ray data have established no evidence of an accretion column, and spectral analysis of the subdominant pulsed component suggests it is non-thermal \citep{Takata2018}. These two characteristics of observed spin-down and absence of an accretion disk led \cite{Buckley2017} to attribute the observed non-thermal luminosity to magnetic dipole radiation from the WD, conclusively establishing AR Sco as the first known WD pulsar, analogous to rotation-powered neutron star pulsars. 

The WD is highly magnetized (with a { polar surface field} of $B_{\rm p}\sim8\times10^8$~G estimated by setting $\dot{E}_{\rm rot}=\dot{E}_{\rm md}$, with $\dot{E}_{\rm md}$ the energy loss due to a magnetic dipole rotating in vacuum) and its optical emission is strongly linearly polarized (up to$\sim40\%;$ \citealt{Buckley2017}). \cite{2018MNRAS} conducted extensive follow-up optical observations on AR Sco, reporting that the linear flux, circular flux and position angle are coupled to the spin period $P$ of the WD.  The polarization position angle (PPA; $\psi$) vs.\ time indicates clear periodic emission. In this paper, we model the linear polarization signature with the rotating vector model (RVM; \citealt{Radhakrishnan}):
\begin{equation} \label{eq 3.1}
\tan(\psi-\psi_{0})=\frac{\sin\alpha\sin(\phi-\phi_{0})}{\sin\zeta\cos\alpha-\cos\zeta\sin\alpha\cos(\phi-\phi_{0})},
\end{equation}
with $\alpha$ the magnetic inclination angle of the magnetic dipole moment $\bm{\mu}$ and $\zeta$ the observer angle (observer's line of sight), both measured with respect to the { WD's} rotation axis $\mathbf{\Omega}$, $\phi$ the rotational phase (WD spin), and the parameters $\phi_{0}$ and $\psi_{0}$ are used to define a fiducial plane. 
The light curve from \cite{Buckley2017} manifests double peaks with a more intense first peak followed by a dimmer second peak, exhibiting a peak separation of $\sim180^{\circ}$. In addition, the { PPA} makes a $180^{\circ}$ sweep. These facts imply that the WD may be an orthogonal rotator if the emission originates close to its polar caps \citep{Geng2016,Buckley2017}. We show that our solution for $\alpha$ using the RVM { supports} this conjecture.

The observations by \cite{Marsh2016} indicate irradiation of the side of the WD facing the companion. This forms part of the observed sinusoidal radial velocity profile, suggesting that the two stars are tidally locked \citep{Buckley2017, Takata2017, 2018MNRAS}. To account for the observed non-thermal radiation, models invoke injection of relativistic electrons by the companion along the magnetic field lines of the WD, where they are trapped and accelerated \citep{Geng2016,Takata2017,Buckley2017}. However, different proposed scenarios place the non-thermal emission regions in different spatial locales. \cite{Geng2016} noted that the Goldreich-Julian charge number density \citep{Goldreich1969} of the WD is much lower than required by the observed SR spectrum and thus argued that the emission should originate closer to the companion (they suggest at an intrabinary shock caused by the interacting stellar winds; \citealt{Marsh2016,Geng2016}) where the particle number density is higher. They noted that the spin-down luminosity of the WD is sufficient to power the emission of the system. On the other hand, upon an injection of relativistic particles from the companion, emission may originate from near the magnetic poles of the WD by means of pulsar emission mechanisms (with the emission from downward-moving particles being directed toward the WD; cf.\ \citealt{Buckley2017,Takata2017,2018MNRAS,Takata2019}). This is supported by the geometric model of \cite{2018MNRAS} that can reproduce the observed polarization signatures if the emission location is taken to be at the magnetic poles of the WD. In what follows, we demonstrate that the RVM\footnote{ It is important to note that the RVM is a purely geometrical model, and as such cannot make any statements about light curve shapes or spectra expected from AR Sco. Neither can we constrain particle energies or injection rates or the acceleration process within this framework. We defer the construction of a full emission model to a future paper.} provides an excellent fit to the PPA curve, thereby favoring { the magnetospheric scenario, if particles sustain small pitch angles.} However, we note that this hypothesis may not be unique, since \citet{Takata2019} claim that they can also reproduce the polarization properties using an independent emission model { in which the particle pitch angles evolve with time and become quite large at the emission regions, which are incidentally also located relatively close to the WD polar caps.} 

The structure of this article is as follows. In Section~\ref{sec:Method} we discuss the folding of the data, code calibration and fits. We present our results in Section~\ref{Results}. Our discussion and conclusions follow in Section~\ref{sec:Disc} and~\ref{sec:Concl}.

\section{Method}\label{sec:Method}
\subsection{Folding of Data}
We use the PPA data from \cite{Buckley2017}, comprising observations on 14 March 2016 in the $340-900$~nm range. We obtain the minimum PPA of the dataset and define the corresponding time $t_{0}$ as the starting point to fold the dataset. 
A few data points ($\sim 3\%$) deviate from the average PPA vs.\ $\phi$ curve, but we note that the folding is affected by the choice of $t_{0}$. We remedy this convention issue by generating smoothed PPA curves with a Kernel Density Estimation and a Gaussian kernel technique. By inspecting the deviation between the smoothed curve and the folded data, we assign a new convention to the points with a large deviation by shifting these points by $180^{\circ}$ (which is the intrinsic uncertainty of the convention assigned to the PPA, { i.e., there is an ambiguity in the parallel and anti-parallel directions}). We then bin the data into 30 rotational phase bins. The predicted values of $\psi$ from Eq.~(\ref{eq 3.1}) are discontinuous, since the arctangent function is discontinuous at $\phi_{\rm{disc}} = \arccos\left(\tan\zeta/\tan\alpha\right).$
Using this expression, we locate the discontinuities and shift the predicted PPA by $360^{\circ}$ at these phases to finally obtain a smooth, continuous PPA model.
        
\subsection{Code Verification and Best Fit}
We adopt the convention of \cite{Everett2001}, letting $\psi$ increase in the counter-clockwise direction. Thus we  define $\psi' = -\psi$. A Bayesian likelihood approach is employed to constrain the RVM using the optical PPA data. We define our ``best fit'' as the $50^{\rm th}$ percentile or median in the posterior distribution of the model parameters. We verified our code by obtaining and comparing independent RVM fits of radio pulsar data \citep{Everett2001}, yielding consistent $\alpha$ and $\zeta$ values, within uncertainties. 

For the Bayesian analysis, we employ a Markov-Chain Monte Carlo technique \citep{Foreman2013} with 50 walkers, 40,000 steps and a step burn-in of 14,000. We maximize the following likelihood function:
\begin{align} \label{likelihood}
& \ln p(y|\phi,\alpha,\beta,\phi_{0},\psi_{0},f) = \nonumber \\
& -0.5\Sigma_{n}[\frac{(y_{n}-H(\phi,\alpha,\beta,\phi_{0},\psi_{0}))^{2}}{S_{n}^{2}} +\ln(2\pi S_{n}^{2})],
\end{align}
where $S_{n}^{2} = \sigma_{n}^{2} +f^{2}(H(\phi,\alpha,\beta,\phi_{0},\psi_{0}))^{2}$, $y$ represents the data values, $\sigma$ the uncertainties of the data (assumed to follow a Gaussian distribution) and $H$ the model values. The factor $f$ compensates for the case when the PPA uncertainties are underestimated. We assume uniform priors on the cosine of $\alpha$ and $\zeta$ (rather than the angles themselves). For $\cos \zeta$, this is an appropriate choice from Copernican arguments. For $\cos \alpha$, this convention is less justified, but nevertheless ultimately immaterial for the present context; we have verified there is no dramatic change in resulting uncertainties with a uniform prior choice on $\alpha$ rather than $\cos \alpha$. 
 
\section{Results}\label{Results}
\subsection{Constraints on WD Geometry}
The red curve in Figure~\ref{fig:4.1} depicts the best-fit RVM\footnote{ See the Appendix for a parameter study that indicates the behavior of the RVM for different choices of $\alpha$ and $\zeta$.} to the polarization data, with yellow curves associated with a random selection of parameters from the posterior distribution of $\cos\alpha$ and $\cos\zeta$.  
The uncertainties for the best fit are taken to be the $68\%$ probability around the median, i.e., the $16^{\rm{th}}$ and $84^{\rm{th}}$ percentile values. 
From our best fits of these quantities, we obtain $\alpha={86.6^\circ}^{+3.0^\circ}_{-2.8^\circ}$ and $\zeta={60.4^\circ}^{+5.3^\circ}_{-6.0^\circ}$. For the uncertainty parameter $\ln(f)$, only the maximum of $\ln(f)$ is constrained, and $f$ is quite small, thus we conclude the data are described well by the model without the inclusion of this term. 

The correlation or degeneracy seen between $\phi_{0}$ and $\psi_{0}$ in Figure~\ref{fig:4.2} owes to the fact that these parameters translate the model horizontally and vertically; thus, a natural degeneracy exists, since the model is cyclic and a similar fit may be obtained for different choices of $\phi_{0}$ and $\psi_{0}$. This is also the reason why small, disconnected contours could be eliminated from Figure \ref{fig:4.2} by constraining the priors of the nuisance parameters $\phi_{0}$ and $\psi_{0}$. The large duty cycle leads to relatively small uncertainties on both $\alpha$ and $\zeta$ as compared to the case of known radio pulsars. The pulses of the latter typically have small duty cycles and thus relatively large uncertainties on $\alpha$ (the impact angle $\beta=\zeta-\alpha$ is typically better constrained, given visibility requirements, but $\zeta$ may remain ill-constrained).
\begin{figure}[t]
\includegraphics[width=22pc]{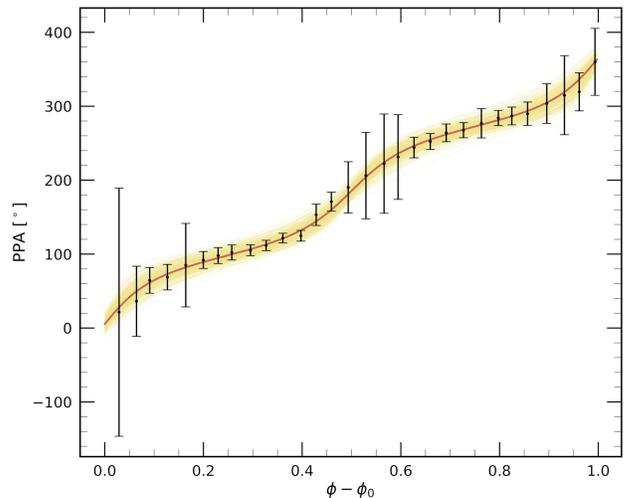} 
\caption{The best-fit RVM solution (red line) for the PPA data using an MCMC technique with the likelihood parameter $f$ included, also showing ensemble plots (possible fits) as yellow lines. { The abscissa is chosen such that the observer crosses the WD fiducial plane, defined by the magnetic moment and spin vectors, at phase zero.}}
\label{fig:4.1}
\end{figure}

\begin{figure}[b]
\includegraphics[width=20pc]{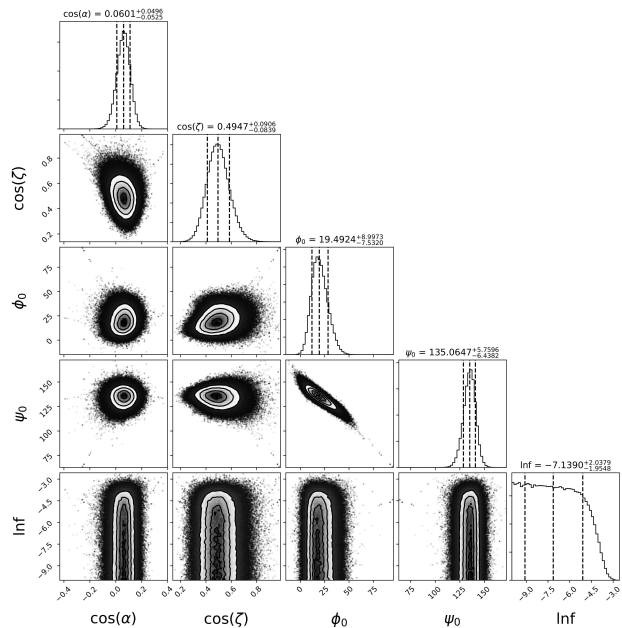} 
\caption{The best-fit for the model implementing $\cos\alpha$, $\cos\zeta$ and a likelihood parameter $f$. We found $\cos\alpha=0.060^{+0.050}_{-0.053}$, $\cos\zeta=0.49^{+0.09}_{-0.08}$, $\phi_{0}=({19.5^\circ}^{+9.1^\circ}_{-8.4^\circ})$, $\psi_{0}=({135.1^\circ}^{+5.8^\circ}_{-6.4^\circ})$, $\ln f=-7.1^{+2.0}_{-2.0}$. The $\phi_{0}$ and $\psi_{0}$ are nuisance parameters.}
\label{fig:4.2}
\end{figure}

\subsection{Constraints on WD Mass}
 \cite{Marsh2016} reported the mass ratio of AR Sco as $1/q = M_{\rm M}/M_{\rm WD} \geq 0.35$, assuming $M_{\rm WD}=0.8M_\sun$ and $M_{\rm M}=0.3M_\sun$ as a natural pairing for the system that is located at a distance of $d=116 \pm 16$ pc. They also measured the radial velocity of the M star, $K_2 = 295 \pm 4$ km s${}^{-1}$ and obtained the following mass function
\begin{align}
m_{\rm WD} \sin^3 i  \left(\frac{q}{1+q} \right)^2 = \frac{P_b K_2^3}{2 \pi G} \approx 0.395^{+0.0163}_{-0.0158} M_\odot.
\label{massfunc}
\end{align}
They report no significant evidence of non-sinusoidal radial velocities, thus suggesting little overestimation of the true amplitude owing to irradiation. Based on the M star's emission line velocities, \cite{Marsh2016} infer $q \lesssim 2.86$, perhaps close to the cited maximum ($q \sim 2.8$) under the assumption that the M star is close to filling its Roche lobe. Moreover, from spectroscopic comparison to model atmospheres they conclude that the radius of the M star $R_{\rm M} \approx 0.36 R_\odot$  using a mass\footnote{ This choice of mass is due to the M5 spectral type of the M-dwarf and is also typical of donor-star masses in other systems having similar orbital periods as AR Sco.} of $m_{\rm M} \approx 0.29 M_\odot$ via the volume-equivalent Roche radius approximation of \cite{1971ARA&A...9..183P},
\begin{align}
\frac{1}{3} \left( \frac{2 G \, m_{\rm M} P_b^2}{3 \pi^2 } \right)^{1/3} \approx R_{\rm M} \approx 0.36 R_\odot.
\label{rocheconstraint}
\end{align}
Finally, based on the stellar radius estimate and the stellar brightness, they quote a mass estimate of $m_{\rm M} \approx  0.3 (d/116 \rm \,  pc)^3 M_\odot$. A more accurate parallax distance of $d=117.8 \pm 0.6$ pc was reported by \cite{Stiller2018}, implying $m_{\rm M} \approx 0.31 M_\odot$ when inverting the above mass-distance relation, suggesting general consistency with the estimate $m_{\rm M} \sim 0.3 M_\odot$.  Substitution of $m_{\rm M} \sim 0.3 M_\odot$ into Eq.~(\ref{massfunc}) and demanding that $m_{\rm WD}$ is below the Chandrasekhar mass limit of $1.44 M_\odot$ yields the constraint $i \gtrsim 47 ^\circ$.

\begin{figure}[t]
\centering
\includegraphics[width=0.47\textwidth]{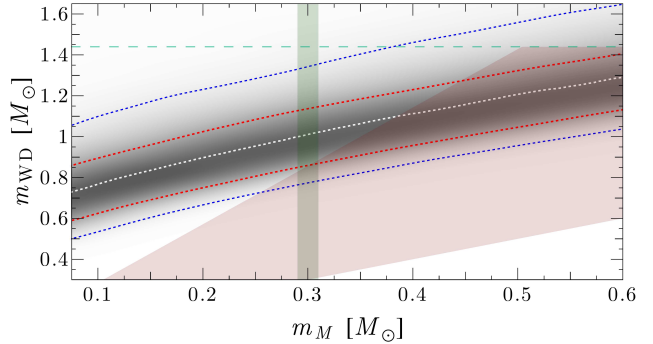}
\caption{Various constraints on the  the component masses $\{m_{\rm M},m_{\rm WD}\}$. The gray band depicts the probability density in the mass function Eq.~(\ref{massfunc}) with the assumption $\zeta \approx i$ constructed by propagating uncertainties from the fitted $\cos \zeta$ in this paper and uncertainties of the radial velocity in \cite{Marsh2016}. The white dashed curve is the median of the probability distribution of $m_{\rm WD}$ while $\{\rm red, blue \}$ curves delineate regions of $\{68\%, 95\%\}$ probability. The pink band is the constraint $1 < q < 2.86$ from \cite{Marsh2016}, while the green band portrays $m_{\rm M} \sim 0.3 M_\odot$ as inferred from Eq.~(\ref{rocheconstraint}). The Chandrasekhar limit of $m_{\rm WD} < 1.44 M_\odot$ is traced by the cyan dashed line. }
\label{massdist}
\end{figure}

The mass $m_{\rm WD}$ is generally\footnote{This is true as long as $q$ is in fact $\gg 1$.  If $q\sim1$ then $m_{\rm WD}$ is sensitive on the value of $q$. For our assumed value of $q\sim3$, there is a small influence of $m_{\rm WD}$.} insensitive to the value of $m_{\rm M}$, as is readily apparent from the $q\gg1$ limit of Eq.~(\ref{massfunc}), and therefore also to any systematic errors associated with observational determinations of $R_{\rm M}$ and also the approximation employed in Eq.~(\ref{rocheconstraint}). Constraints on $m_{\rm WD}$ may be obtained under the assumption $\zeta \sim i$, which is generally expected from formation/evolution and observed in other contexts \citep[e.g.,][]{2007A&A...474..565A,2011MNRAS.413L..71W}. 
Note that the spin angular momentum of the WD is about two orders of magnitude inferior to the total orbital angular momentum. Since it is likely that the WD was spun up to its exceptionally low period of $P=117~$s via past accretion episodes, the transfer of angular momentum would tend to align the WD spin to the orbit angular momentum. Allowing $m_{\rm M}$ be a free parameter, while propagating normally-distributed uncertainties for $K_2$ and $\cos \zeta$ (see Figure~\ref{fig:4.2}) in Eq.~(\ref{massfunc}) yields a band of allowable values of $\{m_{\rm M}, m_{\rm WD} \}$ depicted in Figure~\ref{massdist}. Adopting $R_{\rm M} \approx 0.36 R_\odot$ and solving the system of equations Eq.~(\ref{massfunc}) -- (\ref{rocheconstraint}) yields in $m_{\rm WD} = 1.00^{+0.16}_{-0.10} M_\odot$  for the median, with uncertainties for the $68\%$ containment region of probability. This yields $q=3.45^{+0.66}_{-0.45}$ that is somewhat in tension with \cite{Marsh2016}'s quoted $q \lesssim 2.86$ derived from the velocity amplitudes of atomic emission lines relative to the M, but not meaningfully so, given the allowable range of uncertainties and systematics which may be present in the M star mass/radius estimate. 

\subsection{Geodetic Precession}
If we allow misalignment of the spin and orbital axes $\zeta \neq i$, this may explain why the optical maximum does not occur at inferior conjunction of the WD \citep{Buckley2017}.
{ \citet{Katz2017} proposed that this mismatch may be explained either by dissipation in a bow wave or by misalignment between the WD spin axis and both the orbital axis and the WD's oblique magnetic moment as well as potential oblateness of the WD, causing precession of the spin axis. The latter would lead to variable heating of the companion surface along with a drift in the phase of the optical maximum, with a period of decades. \citet{Peterson2019} analyzed a century's worth of optical photometry on AR Sco but could not detect any precessional period as suggested by \citet{Katz2017}, although this period may be much longer for a larger angle of misalignment between the orbital plane and the WD spin axis, or a smaller oblateness.}

We suggest that effective spin-orbit interaction may result in the precession of the WD pulsar spin axis \citep{1974CRASM.279..971B,1975ApJ...196L...1E} owing to the metric curvature of the companion, { even if the WD is not oblate.} While any precession in the system may be dominated by electromagnetic influences of the companion, the General Relativity (GR) rate calculated below is the minimum rate in absence of any electromagnetic torques.
In the framework of classical GR, the angular geodetic precession rate $\omega_p$ is derived by \cite{1975PhRvD..12..329B,1975ApJ...199L..25B}. The rate of precession is insensitive to $m_{\rm WD}$ and may be estimated independent of the assumption $\zeta \approx i$, 
\begin{align}
\omega_p = & \, \,  \frac{G^{2/3}}{2 c^2 (1- \varepsilon^2)} \left( \frac{2 \pi}{P_b} \right)^{5/3} m_{\rm M}^{2/3} \frac{3 +4 q}{(1+q)^{4/3}} \nonumber \\ 
\sim & \, \, \frac{4 \times 10^{-10}  {\rm \, \, rad \, \, s^{-1}} }{1- \varepsilon^2} \approx  \frac{0.8 {\rm \, \ deg \, \, year^{-1}}}{1- \varepsilon^2}. \label{omegaP}
\end{align}
Since the orbital eccentricity $\varepsilon$ is presumably close to zero, the numerical value in Eq.~(\ref{omegaP}) constitutes a lower limit to the expected precession rate. This rate is similar to pulsar systems where such precession has been detected \citep[e.g.,  B1913+16][]{1989ApJ...347.1030W,1990ApJ...349..546C,1998ApJ...509..856K} over decade timescales. The observable scope of precession depends on the degree of misalignment; if $\zeta$ is significantly different from $i$, such precession (including that by oblateness of the WD) may be imprinted on the pulses and PPA swings of AR Sco and therefore changes in $\zeta$ values may be detectable on similar decade timescales as in pulsar binary systems. That is, from long-term time evolution of pulses and polarization data, the degree of misalignment may be estimated. Moreover, if well-characterized, such precession also affords an independent constraint on the component masses. 

{ \citet{Takata2019} also note that a detailed comparison between model and measured PPA sweeps may constrain the orientation of the spin axis of the WD. 
In future, we will study polarization data as a function of orbital phase $\phi_{\rm b}$ to constrain the effects of precession that a varying $\alpha$ or $\zeta$ vs.\ $\phi_{\rm b}$ may point to. In addition, predictions from a full emission model should lead to predictions of the PPA evolution that may be fit to data to constrain the WD spin axis alignment with the orbital axis and / or precession in the system. It is hoped that contraints on precession may teach us more about the system's characteristics, analogous to the case of PSR J1906+0746 where observations over several years of this precessing pulsar revealed the average structure of the radio beam \citep{Desvignes19}.}

\section{Discussion}\label{sec:Disc}
\subsection{Assumptions Regarding the Particle Pitch Angle}
Using our fits for $\alpha$ and $\zeta$, we can constrain the geometry of the emission region for the optical radiation. The derivation of the RVM \citep[see Appendix of][]{HEASA2019} assumes that the observer samples emission that is \textit{tangent} to local magnetic field lines and that the polarization vector is in the poloidal plane. This is a good approximation in either SR or curvature radiation (CR) scenarios, provided that the particle momentum parallel to the field is relativistic and the relativistic particles have small pitch angles.
 
{ At first sight, this assumption of small pitch angle $\eta$ seems plausible since, for a large Lorentz factor $\gamma$, the pitch angle is $\eta\sim\theta_\gamma \sim 1/\gamma$ (see Eq.~(\ref{eq:gam_min}) where we estimate that the particles radiating optical emission have $\gamma\sim30-100$ from the spectrum, depending on $B$ and $\eta$, and that $\eta\sim10^{-4}$ if $\gamma\sim5$). However, both $\gamma$ and $\eta$ evolve with distance as the particles move along the $B$-field lines. For example, \citet{Takata2017} solve an approximate form\footnote{ The set of equations used by \citet{Takata2017} are valid only for $\gamma \gg 1$ and $\eta \ll 1$, and assumes that the accelerating $E$-field is screened. The more general form of the equations used by \citet{HUM05} includes a term involving a non-zero $E$-field, but are subject to the drift approximation, where the motion of the guiding center is considered and the helical motion is averaged over gyrophase. We concur with \citet{Takata2017} that the local accelerating $E$-field (parallel to the local $B$-field) is probably screened -- see Eq.~(\ref{eq:GJ}) below. We do not address the details of the acceleration process in this paper, other than to note it seems sufficient to accelerate particles to very large $\gamma_{\rm e,max}\sim e B R_{\rm comp}/m_{\rm e}c^2 \sim 10^{6}-10^{8}$. Such high factors are needed if SR is to account for the observed X-ray emission, as noted in Eq.~(\ref{eq:gamma_max}).} of the coupled set of equations that describe the evolution of $\gamma$ and perpendicular momentum $p_\perp$, used earlier by \citet{HUM05} in the context of neutron star pulsars. \cite{Takata2017,Takata2018,Takata2019} assume that relativistic particles are injected from the companion and travel into the WD magnetosphere (along closed $B$-field lines, an assumption based on $\alpha\sim60^\circ<90^\circ$) before radiating significantly, since the SR timescale is much longer than the light crossing timescale $r/c$ at the companion position ($r\sim a$). They then study a magnetic mirror effect in which the first adiabatic invariant $\mu \propto p_\perp^2/B$ is conserved and find that if the initial pitch angle is large enough ($\sin\eta_0 > 0.05$, i.e. outside a loss cone set by initial conditions), the mirror effect operates and the pitch angles increase to $\sim\pi/2$ at the mirror point (thus violating the RVM assumption of tangential emission) as the particles move closer to the WD surface. In this case, significant emission occurs at the mirror point, since SR losses increase rapidly with $B$ and $\eta$, before the particles return outward. 

One can solve for the emission height at which the SR loss timescale equals the light crossing timescale $r/c$:
\begin{align}
\frac{r}{a} \approx 0.21\,\gamma_{50}^{1/5}\mu_{35}^{2/5}\eta_{0.1}^{2/5}\sim0.03R_{\rm LC},
\label{pitch}
\end{align}
with $\gamma_{50} = \gamma/50$, $\mu_{35} = \mu/10^{35}$~G\,cm$^{3}$, $\mu=0.5B_{\rm s}R_{\rm WD}^3$ the magnetic moment, and $\eta_{0.1} = \eta/0.1$ the pitch angle. According to the calculations of \citet{Takata2017}, the mirror effect operates and the particles turn around at $r\sim0.25a$, for $\gamma=50$, $\mu = 6.5\times10^{34}$~G\,cm$^3$ and $\eta=0.1$. However, if we choose plausible values of $\gamma=150$ and $\mu=2\times10^{35}$~G\,cm$^3$ (which is closer to that implied by the estimated surface magnetic field $B_{\rm s}$, see Eq.~[\ref{eq:Bs}]), then Eq.~(\ref{pitch}) implies that $r\sim0.34a>0.25a$. This suggests $p_\perp^2/B$ is no longer an invariant and the mirror effect will not operate, but the particles will lose all their energy abruptly suffering catastrophic SR losses. Thus, the pitch angle may never attain very large values.

We thus note that small pitch angles (as assumed by the RVM) are plausible in two cases: (i) There exist some parameter choices where $t_{\rm SR} > t_{\rm cross}$ at a height $r$ that is some substantial fraction of $a$, so that the particles will radiate all their energy before undergoing magnetic mirroring; 
(ii) particles with small enough initial pitch angles (e.g., $\sin\eta_0\lesssim 0.05$) will fall in the loss cone, and will not be impacted by the magnetic mirror.
A complete model, solving the particle dynamics generally (for any $\eta$ and $\gamma$, and not in the drift approximation) to predict the light curves, spectrum and polarization properties of AR Sco will be able to address this issue more fully.}
\subsection{Constraints on the Emission Geometry}
\begin{figure}
  \includegraphics[width=0.45\textwidth]{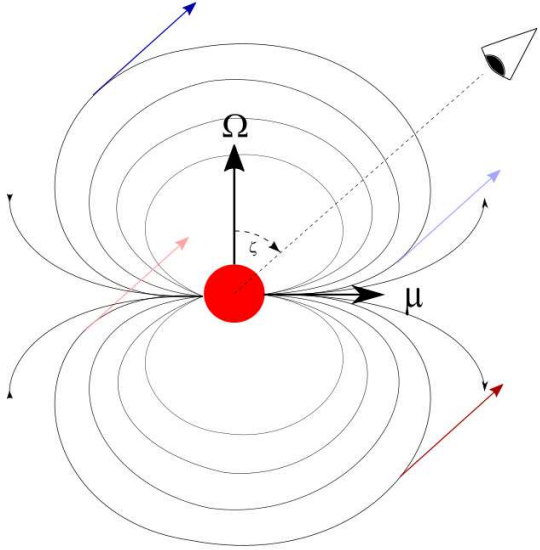}  
  \caption{A schematic diagram indicating an orthogonal rotator with a dipolar field, as well as four possible solutions for emission that is radiated tangent to the local magnetic field lines and is pointing toward a distant observer. The spin axis is indicated by $\bm{\Omega}$, the magnetic moment by $\bm{\mu}$, and the observer angle by $\zeta$. The red arrows indicate inflowing while the blue arrows indicate outflowing particles, as defined with respect to the nearest magnetic pole.}
  \label{cartoon}
\end{figure}
\begin{figure}
 \includegraphics[width=0.45\textwidth]{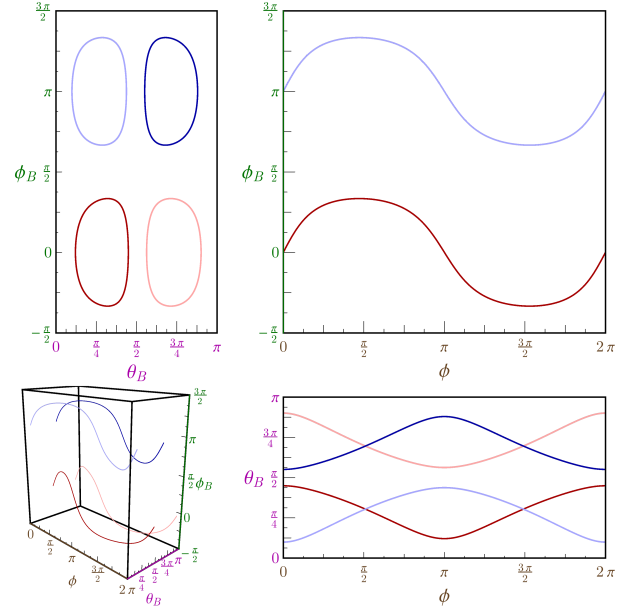}  
 \caption{Magnetic co-latitudes $\theta_{\rm B}$ and longitudes $\phi_{\rm B}$ (with respect to $\bm{\mu}$) which the observer samples during one full rotation in WD spin phase $\phi$ for a static dipole magnetosphere. As in Figure~\ref{cartoon}, the red and blue colors denote inflowing and outflowing charges in a particular magnetic hemisphere (as defined by $\phi_{\rm B}$), defined with respect to the closest magnetic pole. The magnetic poles are located at $\theta_{\rm B} = 0, \pi$.}
 \label{contour}
\end{figure}
In a dipolar magnetosphere, multiple locations satisfy the viewing constraints derivable using our $\alpha$ and $\zeta$ fits { within the framework of the RVM, which assumes emission to be tangential to the local $B$-field}, irrespective of altitude, since a dipolar field is self-similar (additional spectral information may help constrain the actual emission heights, not just the emission directions). { Depending on the current system, any non-zero subset of these locations could participate to produce the observed PPA curve. Thus, since the RVM variation is seen over entire spin rotation of the WD, at least one of the multiple solutions must be realized (e.g., the model \citealt{2018MNRAS} that only assumes inflowing particles would satisfy these constraints).} We indicate this schematically in Figure~\ref{cartoon}, where we show an orthogonal rotator and four different tangents pointing in the observer direction for a particular $\zeta$. Outflowing and inflowing particles with respect to the nearest magnetic pole are indicated by blue and red arrows, respectively. Blue arrows thus occur in the top half of the meridional slice (large magnetic longitude), and red ones in the bottom (small magnetic longitude). Dark and light colors are used to further distinguish between solutions. The meaning of colors remains identical in Figure~\ref{contour}, where we indicate constraints involving the magnetic co-latitude $\theta_{\rm B}$ and longitude $\phi_{\rm B}$ (both defined with respect to the magnetic dipole moment axis $\bm{\mu}$). These constraints derive from the condition that the magnetic field tangents are sampled by the observer, for $\alpha = 87^\circ$ and $ \zeta = 60^\circ$. For $\theta_{\rm B}$, the solutions satisfy Eq.~(34) and its reflection given in \cite{Wadiasingh2018}. 

The different panels in Figure~\ref{contour} are 2D projections of the 3D plot in the leftmost corner. The top left panel indicates that the `red solutions' are located around $\phi_{\rm B}\sim 0$ while the blue ones occur at $\phi_{\rm B}\sim \pi$ (i.e., in meridional and antimeriodional planes with respect to $\bm{\mu}$). In the top right panel, the light and dark colored lines coincide, indicating that the observer samples the same $\phi_{\rm B}$ for each color during the course of one rotation of the WD, independent of $\theta_{\rm B}$. The red and blue solutions are of similar functional form, but offset by a factor $\pi$ in $\phi_{\rm B}$, as previously. The bottom right projection indicates that the light and dark red solutions have the same form (offset by a factor $\sim\pi/2$ in $\theta_{\rm B}$, and the same for the light and dark blue), but the red vs.\ blue ones are mirror images of each other (being $\phi\sim\pi$ out of phase), since they originate close to opposite magnetic poles. Thus, the observer would sample the light blue and dark red solutions (or light red and dark blue ones), offset by half a rotation in spin phase $\phi$, even though they trace out the same range in $\theta_{\rm B}$. 

The constrained spatial region that follows from our RVM fits suggests that a large portion of the magnetosphere, at multiple altitudes, must support relativistic charges. Depending on scenarios of particle acceleration, charges may be either ingoing or outgoing relative to the WD surface, halving the number of potential sites of emission, unless there are counterstreaming beams. In a more complete emission model beyond the geometric RVM, deriving constraints on the altitude of emission should be possible, thereby pinning down the precise location in the magnetosphere where emission arises at any phase.

Interestingly, the estimated polar cap opening angle is much lower ($\theta_{\rm PC}/2\pi\sim1\%$) than the duty cycle $\sim100\%$, implying that the two-star interaction may lead to a much larger opening angle, or that the emission comes from relatively high up, originating on flaring magnetic field lines. 

\subsection{Constraining the Emission Mechanism}
In order to constrain the emission mechanism that might be responsible for the polarized optical radiation, let us estimate several relevant quantities and compare pertinent timescales and lengthscales. The magnetic field at the polar cap may be estimated assuming $M_{\rm WD}=1.0M_{\sun}$ via
\begin{equation}
\begin{split}
B_{\rm p} &\sim 8\times10^8~\rm{G} \\ & \hspace{-5mm} \times\left(\frac{R_{\rm WD}}{7\times10^{8}~\rm{cm}}\right)^{-3}\left(\frac{P}{117~\rm{s}}\right)^{1/2}\left(\frac{\dot{P}}{3.9\times10^{-13}~\rm{s\, s^{-1}}}\right)^{1/2}\label{eq:Bs}\!\!\!\!\!\!\!,
\end{split}
\end{equation} 
from dipole spin-down, with $R_{\rm WD}$ the WD radius.
\cite{Geng2016} calculates the magnetic field strength $B_{\rm x}$ at a distance $x = a-R_{\rm WD}\sim 5\times10^{10}~\rm{cm}$ above the stellar surface, where $a$ is binary separation, obtaining $B_{\rm x}=B_{\rm p}(x/R_{\rm WD})^{-3}\sim 2\times10^3~\rm{G}$.
We estimate the potential drop at the polar cap using \citep{Goldreich1969} 
\begin{equation}
\begin{split}
\Phi_{\rm max} &\sim  4\times10^{11}~\rm{statvolt}\\
& \hspace{-1cm} \times  \left(\frac{R_{\rm WD}}{7\times10^{8}~\rm{cm}}\right)^{3}\left(\frac{B_{\rm p}}{8\times10^8~\rm{G}}\right)\left(\frac{P}{117~\rm{s}}\right)^{-2}\!\!\!\!\!\!\!.
\end{split}
\end{equation}
This yields a maximum Lorentz factor of $\gamma_{\rm{max,\Phi}} \sim 3\times10^{8}$, although it is unlikely that the particle will be able to tap the full potential. 

\begin{figure}[t]
\includegraphics[width=20pc]{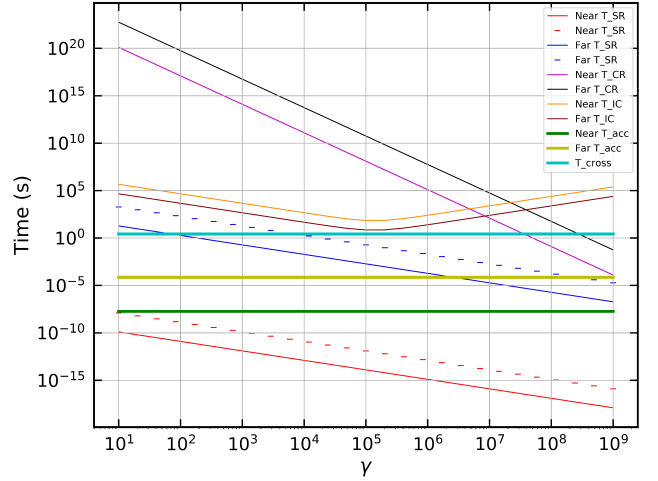} 
\caption{Acceleration, crossing and radiation (CR, SR and IC) loss timescales for `far' ($B_{\rm x}$) and `near' ($B_{\rm p}$) cases. The solid lines for SR represent a pitch angle of $\pi/2$ and the dashed lines represent a pitch angle of $0.1~(\sim6^{\circ})$. { For this figure, we used a companion temperature of $kT=1$~eV, i.e., $T\sim12,000$~K and radius of $R_{\rm comp}=0.36R_\odot$.}}
\label{Time}
\end{figure}
The corresponding SR, CR and inverse Compton (IC) timescales ($\gamma/\dot{\gamma}$) are indicated in Figure~\ref{Time}, along with an acceleration ($R_{\rm L}/c$) and crossing timescale ($a/c$), where $R_{\rm L}$ is the Larmor radius. To calculate the IC loss rate we used $\dot{E}_{\rm IC}=4\sigma_{\rm T}cU\gamma_{\rm e}^{2}\gamma_{\rm KN}^{2}/3(\gamma_{\rm e}^{2}+\gamma_{\rm KN}^{2})$~\citep{Schlickeiser2010}, where $\gamma_{\rm KN}=3\sqrt{5}m_{\rm e}c^{2}/8\pi kT$ and $U=2\sigma_{\rm SB}T^{4}R^{2}_{\star}/cR_{0}^{2}$ is the photon energy density. Here, $m_{\rm e}$ is the electron mass, $\sigma_{\rm T}$ the Thomson cross section, $\sigma_{\rm SB}$ the Stefan-Boltzmann constant, $k$ the Boltzmann constant, $R_{\star}$ the companion radius and $R_{0}$ is the distance from the companion's center to the shock. We consider a `near' ($B_{\rm p}$) and `far' ($B_{\rm x}$) case. Since the IC cooling time is much larger than that of SR (even for small pitch angles), we can rule out IC.

If we equate the SR and CR loss rates to solve for the average pitch angle $\eta$, this yields a maximum value of
\begin{equation}
\begin{split}
\sin\eta 
&\sim 4\times10^{-3}\left(\frac{\gamma}{3\times10^{8}}\right)\left(\frac{B}{2\times10^{3}~\rm G}\right)^{-1}\!\!\!\!\!\!\!\!,
\end{split}
\end{equation} 
where $e$ is electron charge and $\rho_{\rm c}\sim4R_{\rm WD}/3\theta_{\rm pc}$ the curvature radius at the surface of the WD, 
with $\theta_{\rm PC}\sim(\Omega R_{\rm WD}/c)^{1/2}$. \textit{Thus, we see that SR dominates for all reasonable values of $\gamma$ and for non-zero pitch angles, with CR only becoming relevant for $\gamma\gtrsim\gamma_{\rm max,\Phi}$ or for pitch angles $\eta\lesssim10^{-3}$.}

Following \citet{Takata2018} and attributing the non-thermal pulsed X-ray emission to a power-law tail that is emitted by a power-law particle spectrum between $\gamma_{\rm min}$ and $\gamma_{\rm max}$, we can infer constraints using the observed frequencies that correspond to the peak in the spectral energy density ($\nu_{\rm obs,min} = 2\times10^{13}~{\rm Hz}$) and the highest pulsed X-ray photon ($\nu_{\rm obs,max} \gtrsim 2\times10^{18}~{\rm Hz}$):
\begin{eqnarray}
  \gamma^2_{\rm min} B\sin\eta \sim 5\times10^6~{\rm G}\left(\frac{\nu_{\rm obs,min}}{2\times10^{13}~{\rm Hz}}\right),\label{eq:gam_min}\\
  \gamma^2_{\rm max} B\sin\eta \gtrsim 5\times10^{11}~{\rm G}\left(\frac{\nu_{\rm obs,max}}{2\times10^{18}~{\rm Hz}}\right).
\end{eqnarray}
When enforcing that $B<B_{\rm p}$, we obtain
\begin{eqnarray}
  \gamma_{\rm min}  \gtrsim 30\left(\frac{\nu_{\rm obs,min}}{2\times10^{13}~{\rm Hz}}\right)\left(\frac{\sin\eta}{10^{-5}}\right)^{-1/2},\label{gmin}\\
  \gamma_{\rm max}  \gtrsim 8\times10^3\left(\frac{\nu_{\rm obs,max}}{2\times10^{18}~{\rm Hz}}\right)\left(\frac{\sin\eta}{10^{-5}}\right)^{-1/2}\label{eq:gamma_max}\!\!\!\!\!\!\!\!\!\!\!\!,
\end{eqnarray}
for a fiducial value of $\sin\eta\sim10^{-5}$. { (Imposing a lower limit of $\gamma_{\rm min}\approx5$ formally constrains $\eta\gtrsim 2\times10^{-4}$.) These constraints are consistent with the assumptions made by \citet{Takata2017} who model the SR spectrum assuming $\gamma_{\rm min}=50$ and $\gamma_{\rm max}=5\times10^6$.}

Let us calculate a characteristic length scale for both SR and CR:
\begin{equation} \label{LSR}
\begin{split}
L_{\rm SR} &\approx c\frac{\gamma}{\dot{\gamma}_{\rm SR}} =  
10^{10}~\rm cm~\times\\
 &\left(\frac{\nu_{\rm obs,min}}{2\times10^{13}~\rm Hz}\right)^{-1/2}\left(\frac{\rm B}{8\times10^{8}~\rm G}\frac{\sin\eta}{10^{-5}}\right)^{-3/2}
\end{split}
\end{equation}
and
\begin{eqnarray} \label{LCR}
&& L_{\rm CR}  = c\frac{\gamma}{\dot{\gamma}_{\rm CR}} \nonumber \\ 
&& = 10^{20}~{\rm cm}
\left(\frac{\nu_{\rm obs,min}}{2\times10^{13}~\rm Hz}\right)^{-1}\left(\frac{\rm \rho_{\rm c}}{5\times10^{10}~\rm cm}\right).
\end{eqnarray}
Since $L_{\rm SR}\ll L_{\rm CR}$, it is apparent that SR should dominate over CR, even for very small values for $\eta$ (cf.\ Figure~\ref{Time}).

{ Observations by \citet{Buckley2017} indicate that the degree of linear polarization of the optical data varies with orbital phase, but may reach up to 40\%. This constraint should be exploited in an emission model, but is beyond the RVM's capabilities. For example, \citet{Takata2019} argue that a polarization degree that varies with orbital phase may be understood in the framework of different contributions of the SR and the thermal emission from the companion, since the latter may depolarize the observed emission. We expect that a model that includes emission from different emission heights that bunch in phase to form the peaks will not likely overpredict the maximum observed polarization degree of $40\%$. SR from a single particle in an ordered field may reach high values, but this will be lowered by contributions from several particles within the population of radiating particles emitting at different spatial locales. This is similar to the findings of \citet{Harding2017} who found large PA swings and deep depolarization dips during the light-curve peaks in all energy bands in their multiwavelength pulsar model that invoke caustic emission. Such a model may not be exactly applicable to the WD pulsar, although contributions from emitting particles at different altitudes may provide a blended polarization signature, thus lowering the polarization degree vs.\ that expected from a single particle in an ordered field. The detailed characteristics depend on emission position and mechanism.}

A constraint on $B$ can be obtained by assuming that the cyclotron energy is below the SED peak (otherwise a break would be observable due to this threshold energy). At the WD surface, the cyclotron energy is 9~eV. When $\nu_{\rm cycl} = \nu_{\rm obs,min}$, $B \lesssim 7\times10^6$~G, implying that the emission site is at least $\sim5R_{\rm WD}$ above the stellar surface. Another constraint on $B$ may be obtained by requiring the ratio of electromagnetic to kinetic particle energy density to be $\sigma\gg 1$ (since the RVM results point to particles following the local field), but this limit is not very constraining due to the uncertain emission volume. On the other hand, the plasma density should be high enough to explain the observed flux of non-thermal emission.  To estimate a lower limit for the plasma density (since the charge density may be much lower than the actual number density) we can calculate the Goldreich-Julian number density $n_{\rm GJ}$ \citep{Goldreich1969}:
\begin{equation}
\begin{split}
|n_{\rm GJ}| &= \frac{\bold{\Omega}\cdot\bold{B}}{2\pi ec}\\ &= 4.7\times10^5 ~\rm{cm^{-3}}\left(\frac{P}{117 ~\rm s} \right)^{-1}\left(\frac{B}{8\times10^8 ~\rm G} \right) \\ &= 1.2 ~\rm{cm^{-3}}\left(\frac{P}{117 ~\rm s} \right)^{-1}\left(\frac{B}{2\times10^3 ~\rm G} \right),\label{eq:GJ}
\end{split}
\end{equation}
with $\bold{\Omega}$ the angular frequency. 
{ The number of particles needed to explain the SR spectrum may be estimated as follows. Let us focus on the peak frequency $\nu_{\rm peak} = 0.29\nu_{\rm c}$, with $\nu_{\rm c} = 3e\gamma^2 B\sin\eta/4\pi m_{\rm e} c$. We use $F(\nu_{\rm peak}/\nu_{\rm c}) = F(0.29) = 0.924$ and assume a delta distribution for the steady-state particle spectrum $dN/dE_{\rm e} = N_0\delta(E_{\rm e} - E_{\rm e}^*)$, with $E_{\rm e}^* = \gamma m_{\rm e}c^2$ the energy needed to explain the peak of the spectrum at $0.29\nu_{\rm c}$. From Figure 5 of \citet{Takata2019}, we take $\nu F_\nu^{\rm obs} = 8\times10^{-12}$~erg/s/cm$^2$ at an observed energy of $\sim 0.02$~eV. 
Using
\begin{equation}
P_\nu = \frac{\sqrt{3} e^3 B\sin\eta}{m_{\rm e}c^2}F(\nu_{\rm peak}/\nu_{\rm c})
\end{equation}
and a rough estimate of the emitting volume $V\sim2\pi(1-\cos\theta_{\rm PC})a^3/3\sim10^{30}$~cm$^3$, we find 
\begin{equation}
N_0\sim 5\times10^{36}
\end{equation}
and thus\footnote{ One may use this rough estimate of particle number density as well as assumptions for / limits on $\sigma=B^2/(8\pi n_{\rm e}\gamma m_{\rm e}c^2)$ at different spatial positions, implying different values of $B$, to constrain the average $\gamma$ at those positions. E.g., at the companion position, one would obtain $\langle \gamma \rangle \lesssim  10^4$ for $\sigma>1$.}
\begin{equation}
n_{\rm e}\sim \frac{N_0}{V} \sim10^6~{\rm cm}^{-3}. 
\end{equation}
This is lower than the estimate of \citet{Geng2016} who find $n_{\rm e}\sim4\times10^8$~cm$^{-1}$, but still highly supra-Goldreich-Julian suggesting screening of $E$ fields and electron-ion plasma sourced from the companion.}
On the other hand, since we have demonstrated that the PPA of the WD can be modeled by the RVM, this could mean that the emission of the AR Sco system originates from a dipole-like magnetosphere surrounding the WD \citep{Buckley2017}. Using observations of the linear flux, circular flux and PPA observations, \citet{2018MNRAS} show that the polarization is coupled to the WD spin period, agreeing with our conclusion, while models that place the emission site at the companion fail to explain the polarization signatures that are clearly coupled to the spin period of the WD. The `missing' particles noted by \citet{Geng2016} may either be supplied by the companion that may inject relativistic electrons into the WD magnetosphere, or less likely by pair cascades (probably needing severely non-polar B-field structures) occurring in the WD magnetosphere, and does not \textit{per se} argue for an emission location near a putative bow shock. Moreover, freshly-injected particles from the companion may solve the issue of needing non-zero pitch angles for SR to dominate.

\section{Conclusions}\label{sec:Concl}
We applied the RVM to optical polarization data of AR Sco. We obtained $\alpha \sim 90^{\circ}$ confirming the value expected from light curve inspection and a $180^{\circ}$ PPA swing with spin phase. Finding $\zeta \sim 60^{\circ}$ also agrees with the independent assumption by \cite{2018MNRAS} who adopted $\zeta = 60^{\circ}$ for their model to reproduce the observed data. Using the result of $\zeta = 60^{\circ}$ we then found that $m_{\rm WD} = 1.00^{+0.19}_{-0.13} M_\odot$, which is within the limits by \cite{Marsh2016}. We next obtained $q = 3.45^{+0.66}_{-0.45}$, which is slightly larger compared to the value of $q=2.67$ calculated by \cite{Marsh2016} when adopting $m_{\rm WD}=0.8$.
{ Our fits of the PPA evolution of the WD using the RVM could imply that the emission of the AR Sco system originates from the dipole-like magnetosphere of the WD, probably close to its polar caps, while the particles are probably being accelerated at and injected from the companion star. The issue of pitch angle evolution should be addressed with detailed emission modelling.}

Future work includes fitting the RVM to PPA data in other energy bands, applying a general polarization calculation that includes special-relativistic corrections and investigating the effect of different types of magnetic field structures on the predicted polarization signatures. We will also model the phase-resolved PPA to infer $\alpha$ and $\zeta$ for different orbital phases, which may constrain effects of spin precession in this system. This work highlights the complementary constraints on system geometry, emission locales and radiation physics that are supplied by adding polarization data to spectral and temporal data within a unified approach.

\acknowledgements
\hfill \break
\noindent 
We thank the anonymous referee for insightful questions, comments and suggestions. Z.W.\ thanks Demos Kazanas for helpful discussions. This work is based on the research supported wholly / in part by the National Research Foundation of South Africa (NRF; Grant Number 99072). The Grantholder acknowledges that opinions, findings and conclusions or recommendations expressed in any publication generated by the NRF supported research is that of the author(s), and that the NRF accepts no liability whatsoever in this regard. Z.W. is thankful for support from the NASA Postdoctoral Program. This work has made use of the NASA Astrophysics Data System.

\software{ {\tt emcee} \citep{Foreman2013}}

\appendix
\section{RVM Atlas}
\begin{figure}[t]
\begin{center}
\includegraphics[width=0.99\textwidth]{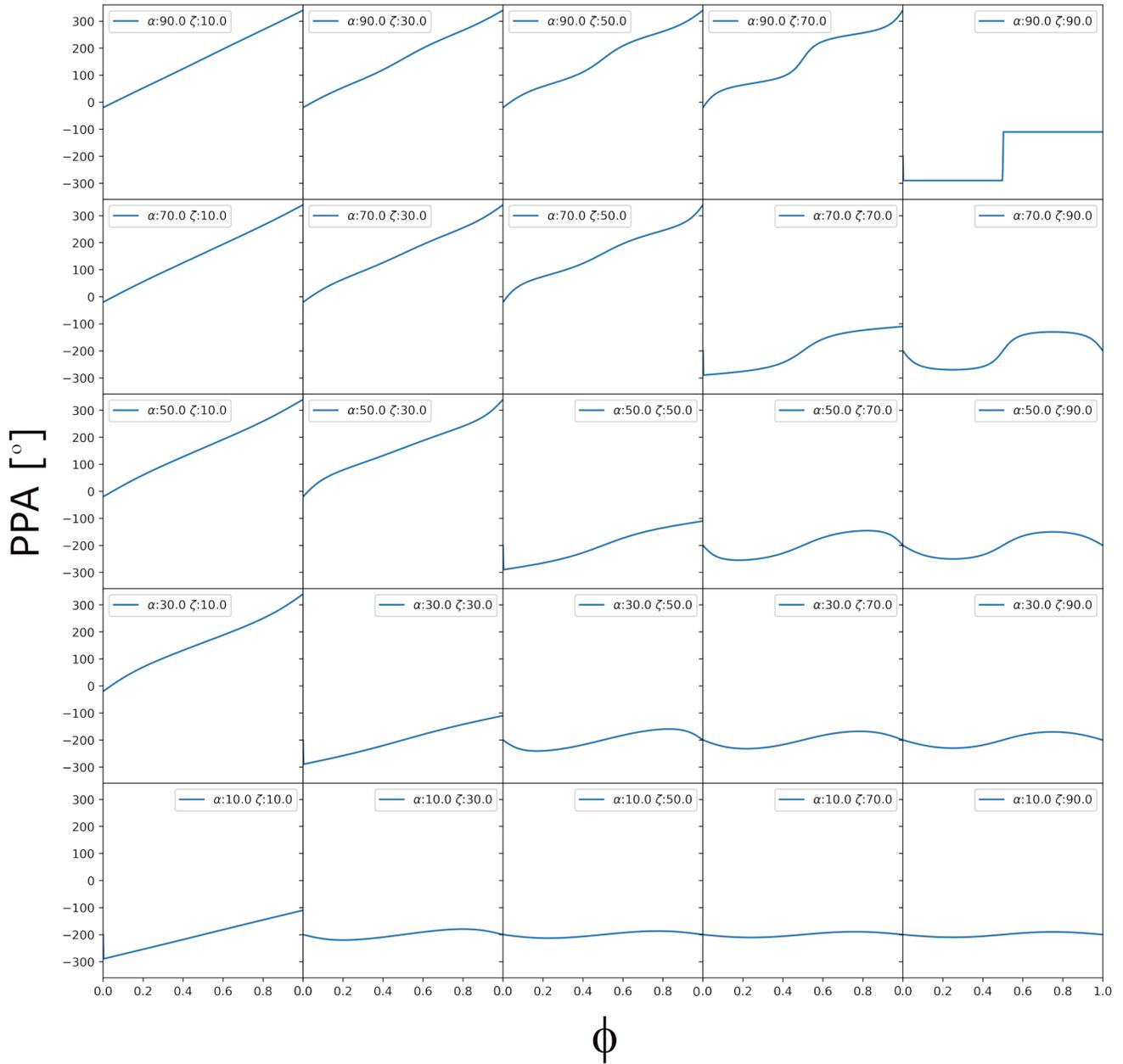} 
\end{center}
\caption{Atlas of RVM for different choices of $\alpha$ and $\zeta$ (in degrees) as indicated in the legend of each subplot.}
\label{RVM_atlas}
\end{figure}

{ In Figure~\ref{RVM_atlas}, we indicate the behavior of the RVM for different choices of $\alpha$ and $\zeta$.}

\bibliographystyle{aasjournal}
\bibliography{ARsco_refs}
\end{document}